\documentstyle[12pt]{article}
\def\deg{\ifmmode{^{\circ}}\else ${^{\circ}}$\fi}

\newcommand{\bi}{\begin{itemize}}
\newcommand{\ei}{\end{itemize}}
\def\ed{\end{document}}
\begin{document}
\newcommand{\gsim}{\,\raisebox{-0.13cm}{$\stackrel{\textstyle
           >}{\textstyle\sim}$}\,}
\newcommand{\lsim}{\,\raisebox{-0.13cm}{$\stackrel{\textstyle
           <}{\textstyle\sim}$}\,}
\newcommand{\lqcd  }{ \mbox{$\Lambda_{QCD} $} }
\newcommand{\lqcdsq  }{ \mbox{$\Lambda^2_{QCD} $} }
\begin{titlepage}
\begin{flushright}
{\sl NUB-3086/94-Th}\\
hep-ph/9403356\\
{\sl March 1994}
\end{flushright}
\vskip 0.5in
\begin{center}
{\Large\bf An alternative model for the electroweak symmetry breaking
sector and its signature in future $e \gamma$
colliders. }\\[.5in]
{\sc Rogerio Rosenfeld}\footnote{e-mail: rosenfeld@neuhep.hex.neu.edu}
\\[.5in]
{\sl Department of Physics}\\
{\sl Northeastern University}\\
{\sl Boston, MA 02115}
\end{center}
\vskip 1in
\begin{abstract}
We perform a preliminary study of the deviations from the Standard Model
prediction for the cross section for the process
$e \gamma \rightarrow \nu_e W \gamma$. We work in the context of a higgsless
chiral lagrangian model that includes an extra vector resonance $V$ and
an anomalous $\gamma W V$ coupling. We find that this cross section
can provide interesting constraints on the free parameters of the model once it
is measured in future $e \gamma$ colliders.
\end{abstract}
\end{titlepage}
\clearpage
\setcounter{page}{2}
\section{Introduction}
\hspace*{\parindent}

The Standard Model has been tested to a remarkable accuracy at LEP
but there is still very little information about its electroweak
symmetry breaking (EWSB) sector apart from a lower limit on the Higgs
boson mass, $M_H > 63.5$ GeV \cite{LEP} .
On the theoretical side, the existence of a Higgs boson
by itself makes it
difficult to understand the fine-tuning that protects its mass to be
as large as the Planck mass.
There are two possible ways to avoid this so-called hierarchy
problem \cite{hierarchy}. One could either have a supersymmetric theory,
in which quantum
corrections to the Higgs mass are only logarithmic instead of quadratic,
or one could have
a strongly interacting underlying theory which has the EWSB sector
as its low-energy effective theory.
In spite of some new indirect evidence in favor of supersymmetric
models \cite{Amaldi}, here we will concentrate on the second approach,
which has rich
experimental consequences at the TeV scale that could be tested at future
hadron colliders.

The presence of this strongly interacting sector would manifest itself
primarily in the scattering of the longitudinal components of the weak
gauge bosons $W_L,Z_L$, which at high energies behave as the pseudo
Nambu-Goldstone bosons originating from the global
symmetry breaking occurring in the EWSB sector,
$SU(2)_L \times SU(2)_R  \rightarrow SU(2)_V $.
The residual $SU(2)_V$ symmetry is responsible for keeping $\rho =
M_W/(M_Z \cos \theta) =  1$ at tree level.
Therefore, $W_L$'s and $Z_L$'s behave as techni-pions of this underlying
strong dynamics and we can describe their interactions by using chiral
lagrangian techniques.
If this scenario is correct, it is very plausible
that resonances would also play a role in describing physical processes.
In fact, amplitudes derived solely from the pion sector of a chiral lagrangian
grow with energy and violate unitarity rather quickly; the presence of
resonances tend to unitarize these amplitudes.
At this point, the simplest possibility is that two types of resonances could
appear: scalar
and vector resonances. The usual Higgs sector is an example of the former
choice; however, in QCD we know that vector resonances are more important in
restoring unitarity, {\it i.e.}, in saturating scattering amplitudes.
Therefore,
we should keep an open mind with respect to what type of resonance would
show up in $W_L W_L$ scattering.
Both types of resonances can be nicely described in terms of chiral
lagrangians.
The possibility of strong $W_L W_L$ interactions has been recently studied in
detail at future hadron colliders \cite{hadron} and at future
$e^+ e^-$ machines \cite{e+e-}.

In this work we will concentrate on a higgsless chiral lagrangian model,
 the so-called BESS model \cite{Casalbuoni}, which
incorporates an
extra vector resonance $V$ analogous to a techni-rho .
We extend this model by including new anomalous terms that give rise to
$\gamma W V $ interactions. We explore the consequences of such an interaction
at future $e  \gamma$
colliders, obtained from a $\sqrt{s} = 1$ TeV $e^+ e^-$ machine via laser
back-scattering off one of the high energy electron (positron) beam.
This $\gamma W V $ coupling is the analogue of the coupling responsible for
$\rho \rightarrow \pi \gamma$ decays in QCD.
In a previous work \cite{rosenfeld} a similar model was used to estimate
the contribution from this new resonance to
$\gamma \gamma \rightarrow Z_L Z_L$ but the results were not encouraging
due to a lack of a $s-$channel contribution.
A similar $\gamma Z \omega_T $ coupling, where $\omega_T$ is the techni-omega
has been used to study the possibility of detecting a $\omega_T$ in future
hadron colliders \cite{chivukula-golden}.

In particular, we will study the sensitivity of the cross section for the
process $e \gamma \rightarrow \nu_e W \gamma$ to a free parameter $k$
that parametrizes this new coupling.
The advantages of studying such a process is that it is independent of
the usual Standard Model Higgs boson and the new vector particle with the
anomalous coupling would appear as a $s-$channel resonance.

This paper is organized as follows : in the next section we review the
salient features of the BESS model; in section 3 we extend the model
by introducing  anomalous couplings terms responsible for the $\gamma W V $
coupling; in section 4 we compute the cross section for
the process $e \gamma \rightarrow
\nu_e W \gamma$  in the presence of the new anomalous coupling and
we conclude in section 5.

\section{Brief review of the BESS model}

\hspace*{\parindent}

The BESS model \cite{Casalbuoni} is based on the so-called hidden symmetry
approach
\cite{Bando} that
successfully describes the low energy QCD phenomenology involving
 $\rho$'s and $\pi$'s. Applied to the electroweak sector,
this hidden symmetry approach describes the interactions of a new vector
resonance $V$ with the particles of a higgsless Standard Model.
In this section we will summarize its relevant
results.

Let us begin by writing down a global $SU(2)_L \times SU(2)_R$ non-linear
$\sigma$ model which describes the
low-energy interactions of the ligh degrees of freedom (the Nambu-Goldstone
bosons $w^a (a = 1,2,3)$) of a strongly
interacting fundamental gauge theory:
\begin{equation}
{\cal L} = \frac{v^2}{4} Tr[ \partial_\mu U \partial^\mu U^\dagger ] ,
\end{equation}
where
\begin{equation}
U = \exp [2 i w/v] \;\;\; , \;\;\; w = w^a \tau^a/2 \;\;\;,\;\;\;
v = 246 \; \mbox{GeV} .
\end{equation}
The above lagrangian is invariant under the transformation
$ U \rightarrow L U R^\dagger$ , where $L,R \in SU(2)$.

The simplest BESS model is obtained by introducing a vector resonance
as a gauge boson of an additional local $SU(2)_V$ symmetry. Furthermore,
we will gauge the   $SU(2)_L \times SU(2)_R$ symmetry to introduce the
usual gauge fields. To this end we re-parametrize $U$ as :
\begin{equation}
U(x) = \xi_L^\dagger (x) \; \xi_R (x) ,
\end{equation}
where under a local $SU(2)_L \times SU(2)_R \times SU(2)_V$ transformation
we have :
\begin{equation}
\xi_L \rightarrow h \xi_L L^\dagger \; \; , \; \;
\xi_R \rightarrow h \xi_L R^\dagger  \; \; , \; \; h \in SU(2)_V  .
\end{equation}
The fields $\xi$ can be represented in the following form :
\begin{equation}
\xi_L = \exp [i \sigma/v] \; \exp[-i w/v]   \; \; , \; \;
\xi_R = \exp [i \sigma/v] \; \exp[i w/v]  \; \; ,
\end{equation}
where $\sigma = \sigma^a \tau^a/2$ is sometimes called a compensator field.

Under a parity transformation in the strongly interacting sector one has :
\begin{equation}
\sigma \leftrightarrow \sigma \; \; , \;\; w \leftrightarrow -w \; \; , \; \;
\xi_L \leftrightarrow \xi_R \; \; , \; \; U \leftrightarrow U^\dagger \; \; ,
\; \;
\vec{x} \leftrightarrow -\vec{x} .
\end{equation}

The gauge fields  $\tilde{V}_\mu$ , $l_\mu$ and $r_\mu$ transform as usual :
\begin{equation}
\tilde{V}_\mu \rightarrow h \tilde{V}_\mu h^{\dagger} + \frac{i}{g_V}
(\partial_\mu h)
h{^\dagger}  \;\;\; ,\;\;\;
l_\mu \rightarrow L l_\mu L^{\dagger} + \frac{i}{g_L} (\partial_\mu L)
L{^\dagger}  \;\;\; ,\;\;\;
r_\mu \rightarrow R r_\mu R^{\dagger} + \frac{i}{g_R} (\partial_\mu R)
R^{\dagger}  .
\end{equation}

The covariant derivatives are given by :
\begin{eqnarray}
D_\mu \xi_L &=& \partial_\mu \xi_L - i g_V \tilde{V}_\mu \xi_L + i g_L \xi_L
l_\mu
                                                           \nonumber \\
D_\mu \xi_R &=& \partial_\mu \xi_R - i g_V \tilde{V}_\mu \xi_R + i g_R \xi_R
r_\mu ,
\end{eqnarray}
which transform as :
\begin{equation}
D_\mu \xi_L  \rightarrow h \; D_\mu \xi_L \; L^\dagger  \; \; , \; \;
D_\mu \xi_R  \rightarrow h \; D_\mu \xi_R \; R^\dagger .
\end{equation}

We finally define the ``building blocks" $\alpha_{L \mu}$
and $\alpha_{R \mu}$ to construct a chiral invariant
lagrangian by :
\begin{equation}
\alpha_{L \mu} = D_\mu \xi_L \; \xi_L^\dagger  \; \; , \; \;
\alpha_{R \mu} = D_\mu \xi_R \; \xi_R^\dagger ,
\end{equation}
which transform under local $ SU(2)_L \times SU(2)_R \times SU(2)_V$ as :
\begin{equation}
\alpha_{L \mu} \rightarrow h \alpha_{L \mu} h^\dagger  \; \; , \; \;
\alpha_{R \mu} \rightarrow h \alpha_{R \mu} h^\dagger .
\end{equation}
Under parity we have  $\alpha_{L 0} \leftrightarrow  - \alpha_{R 0}$ and
 $\alpha_{L i} \leftrightarrow  \alpha_{R i}$ .

Finally , the most general
lowest order lagrangian which respects the parity-like symmetry
$L \leftrightarrow R$ is given by :
\begin{equation}
{\cal L} = -
\frac{v^2}{4} Tr[ ( \alpha_{L \mu} - \alpha_{R \mu} )^2 ] - a \frac{v^2}{4}
 Tr[ ( \alpha_{L \mu} + \alpha_{R \mu} )^2 ]  + {\cal L}_{\mbox{kin}}
\end{equation}
where
\begin{equation}
 {\cal L}_{\mbox{kin}} =  -\frac{1}{2} Tr[ (V_{\mu \nu})^2 ] -
\frac{1}{2} Tr[ (L_{\mu \nu})^2 ]  -\frac{1}{2} Tr[ (R_{\mu \nu})^2 ]
\end{equation}
are the kinetic terms involving the usual non-abelian
field-strength for the vector resonance and gauge fields.

In unitary gauge ($w = \sigma = 0$, $\xi_L = \xi_R = 1$) and specializing
the gauge fields to the Standard Model fields the $\alpha_\mu$'s reduce
to :
\begin{equation}
\alpha_{L \mu} = -i g_V \tilde{V}_\mu + i g \tilde{W}_\mu  \; \; , \; \;
\alpha_{R \mu} = -i g_V \tilde{V}_\mu + i g' B_\mu \tau^3/2 ,
\end{equation}
and the usual BESS model is given by the lagrangian :
\begin{eqnarray}
{\cal L}_{\mbox{BESS}} &=& - \frac{v^2}{4} \left\{
Tr[ (i g \tilde{W}_\mu - i g' B_\mu \tau^3/2)^2 ] + a Tr[ (i g \tilde{W}_\mu +
i g' B_\mu \tau^3/2 - 2 i g_V \tilde{V}_\mu)^2 ] \right\} \nonumber \\
                       & &  +  {\cal L}_{\mbox{kin}}
\label{eq:bess}
\end{eqnarray}

Notice that here we prefer to keep a canonical kinetic term for the
$\tilde{V}_\mu$ field and for this reason our coupling constant $g_V$ differs
from the coupling constant $g''$ used in references \cite{Casalbuoni} :
$g_V = g''/2$ .
Since it is not relevant in what follows, we assume that there is no
direct coupling between the new vector resonance $\tilde{V}_\mu$ and fermions.
This is equivalent to set the parameter $b$ defined in reference
\cite{Casalbuoni} to zero.

The fields appearing in Eq. \ref{eq:bess} are unphysical and one obtains
the physical fields $A_\mu$, $Z_\mu$, $W_\mu$ and $V_\mu$ by diagonalizing
the mass matrix in the neutral and charged sector. In the charged sector one
has :
\begin{eqnarray}
\tilde{W}^\pm &=& W^\pm \; \cos \phi  + V^\pm \; \sin \phi   \\
\tilde{V}^\pm &=& - W^\pm \; \sin \phi +  V^\pm \; \cos \phi .
\end{eqnarray}
In the neutral sector the situation is slightly more complicated :
\begin{eqnarray}
\tilde{W}_3 &=&  A_1 \; \cos \theta + A_2 \; \sin \theta \\
B &=& - A_1 \; \sin \theta +  A_2  \; \cos \theta \\
\tilde{V}_3 &=& V \; \cos \xi \; \cos \psi  - Z \; \sin \xi \; \cos \psi  +
A \; \sin \psi
\end{eqnarray}
where
\begin{eqnarray}
A_1 &=& Z \; \cos \xi  + V \; \sin \xi   \\
A_2 &=& A \; \cos \psi  + Z \; \sin \psi \; \sin \xi  - V \; \sin \psi \;
\cos \xi
\end{eqnarray}

The mixing angles introduced above are given by :
\begin{equation}
\tan \theta = g'/g \; \; , \; \; \tan \psi = \frac{g' \cos \theta}{g_V} =
\frac{g \sin \theta}{g_V}
\end{equation}
and in the limit $g_V \gg g,g'$ :
\begin{equation}
\phi = -\frac{g}{2 g_V}  \; \; , \; \; \xi = - \frac{g^2 - g'^2}{g_V
                                              \sqrt{g^2 + g'^2} } .
\end{equation}

The physical photon field is massless ($M_A = 0$),  and in the
limit $g_V \gg g,g'$ the masses of the other vector bosons are given by :
\begin{eqnarray}
M_W^2 &=& \frac{v^2}{4} g^2 \left( 1 - \left( \frac{g}{2 g_V} \right)^2
                                                                 \right) \\
M_Z^2 &=& \frac{v^2}{4} (g^2+g'^2) \left( 1 -
 \frac{(g^2 - g'^2)^2}{(g^2 + g'^2) 4 g_V^2}   \right) \\
M_V^2 &=& a g_V^2 v^2
\end{eqnarray}

The Standard Model couplings are also modified due to mixing. For instance,
the electric charge in this case is given by : $e = g \; \sin \theta \;
\cos \psi$ .

There are two free parameters in this lagrangian, namely, $g_V$ and $a$.
In the limit $g_V \rightarrow \infty$, one recover the mass relations and
couplings
of the Standard Model. Precision measurements from LEP put constraints
on the parameters of the model \cite{Casalbuoni4}.

The width of the vector resonance (for $M_V \gg M_W$ ) is given by :
\begin{equation}
\Gamma_V = \frac{a M_V^3}{192 \pi v^2}
\end{equation}
and we can substitute ($M_V$, $\Gamma_V$) for ($g_V$, $a$) as free parameters.

In this work we are interested in a possible $\gamma W V$ vertex and
in principle one might think that such a vertex could occur in the kinetic
terms after substituting for the physical fields. However, an explicit
computation shows that no such interaction appears. In fact, one can
show that if such an interaction resulted from the kinetic terms then
electromagnetic gauge invariance would be violated in the process
$\gamma W \rightarrow \gamma W$.
Therefore, we are led to expand the BESS model by introducing extra terms
that induce new ``anomalous" interactions.
In the next section we discuss the motivation
and describe this anomalous BESS model.

\section{ The anomalous BESS model }

\hspace*{\parindent}

The parity-like operation $L \leftrightarrow R$ is not a symmetry
of the underlying theory. It corresponds to a symmetry of the theory under
$w(\vec{x},t) \rightarrow -w(\vec{x},t)$ , which forbids transitions between
even and odd numbers of the pseudo-Goldstone bosons $w$. However, parity
conservation implies in the symmetry $w(\vec{x},t) \rightarrow -w(-\vec{x},t)$
and it is possible to write down parity-conserving terms in the lagrangian
that violate the  $L \leftrightarrow R$ symmetry \cite{anomalous}.
In QCD, these terms describe processes like $\rho, \omega \rightarrow
\pi \gamma $.

In order to write the possible terms down we go back to the
general case and define
the following modification of the field strength for the external
gauge fields  :
\begin{equation}
\hat{L}_{\mu \nu} = \xi_L L_{\mu \nu} \xi^\dagger_L \; \; , \; \;
\hat{R}_{\mu \nu} = \xi_R R_{\mu \nu} \xi^\dagger_R
\end{equation}
so that under $SU(2)_L \times SU(2)_R \times SU(2)_V $ we have
$V_{\mu \nu}$,$\hat{L}_{\mu \nu}$ and $\hat{R}_{\mu \nu}$ transforming
in the same manner as the ``building blocks" $\alpha_\mu$'s.
Under parity we have :
\begin{eqnarray}
V_{00} &\leftrightarrow& V_{00} \; \; , \; \; V_{0i} \leftrightarrow -V_{0i}
\; \; , \; \; V_{ij} \leftrightarrow V_{ij}  \nonumber \\
\hat{L}_{00} &\leftrightarrow& \hat{R}_{00} \; \; , \; \;
\hat{L}_{0i} \leftrightarrow -\hat{R}_{0i}
\; \; , \; \; \hat{L}_{ij} \leftrightarrow \hat{R}_{ij} .
\end{eqnarray}

The extra terms invariant under
$SU(2)_L \times SU(2)_R \times SU(2)_V \times$parity are given by
\cite{anomalous} :
\begin{equation}
{\cal L}_{\mbox{anom}} = \sum_{i=1}^{4} \; \kappa_i \; {\cal L}_i
\end{equation}
where
\begin{eqnarray}
 {\cal L}_1 &=& \epsilon^{\mu \nu \alpha \beta} \;
   Tr[ \alpha_{L \mu} \alpha_{L \nu} \alpha_{L \alpha} \alpha_{R \beta} -
        \alpha_{R \mu} \alpha_{R \nu} \alpha_{R \alpha} \alpha_{L \beta} ] \\
 {\cal L}_2 &=& \epsilon^{\mu \nu \alpha \beta} \;
   Tr[ \alpha_{L \mu} \alpha_{R \nu} \alpha_{L \alpha} \alpha_{R \beta} ] \\
 {\cal L}_3 &=& i \; \epsilon^{\mu \nu \alpha \beta} \;
   Tr[ V_{\mu \nu} ( \alpha_{L \alpha} \alpha_{R \beta} -
                     \alpha_{R \alpha} \alpha_{L \beta} ) ] \\
 {\cal L}_4 &=& i \;\epsilon^{\mu \nu \alpha \beta} \;
   Tr[ \hat{L}_{\mu \nu}  \alpha_{L \alpha} \alpha_{R \beta} -
       \hat{R}_{\mu \nu}  \alpha_{R \alpha} \alpha_{L \beta}  ]
\end{eqnarray}
These terms are formally of order ${\cal{O}}(p^6)$ compared to the
 ${\cal{O}}(p^4)$ terms of Eq. [12] but higher order terms respecting the
$L \leftrightarrow R$ symmetry would not generate the interactions we are
interested in.

In unitary gauge, ${\cal L}_1$ and ${\cal L}_2$ generate only $4-$boson
couplings which are of higher order in the coupling constant. However, no
new $\gamma \gamma W^+ W^-$ coupling is present because of the antisymmetric
tensor. The relevant terms in  ${\cal L}_3$ are naively supressed by a
factor of  $g/g_V$ with respect to  ${\cal L}_4$ and therefore here we
will concentrate on  ${\cal L}_4$.

The relevant terms in  ${\cal L}_4$ are given by, in unitary gauge :
\begin{eqnarray}
 {\cal L}_4 &=& 2i \;  \epsilon^{\mu \nu \alpha \beta} \;
Tr[ \partial_\mu \tilde{W}_\nu \; (-i g_V \tilde{V}_\alpha +
 i g \tilde{W}_\alpha) \; (-i g_V \tilde{V}_\beta + i g' B_\beta \tau_3/2) -
\nonumber  \\
& & \! \! \! \! \partial_\mu B_\nu \tau_3/2 \;  (-i g_V \tilde{V}_\alpha +
i g' B_\alpha \tau_3/2) \; (-i g_V \tilde{V}_\beta +
 i g \tilde{W}_\beta)  ] + {\mbox{$4$-gauge couplings}}
\end{eqnarray}
It is now straightforward but tedious to use the relations Eqs.[$16-24$]
in order
to derive the $\gamma W V$
interaction lagrangian :
\begin{equation}
{\cal L}_{\gamma W V} = -\frac{i}{6} g (\frac{g}{g_V})^2 (\sin \theta -
\cos \theta) \cos \psi \;  \epsilon^{\mu \nu \alpha \beta} \;
\partial_\mu A_\nu \; (V_\alpha^+ W_\beta^- - V_\alpha^- W_\beta^+ )
\end{equation}
Notice that the overall coupling is very small, being of the order of
${\cal O} (g^3/g_V^2)$. However, this term is to be multiplied by a free
parameter $\kappa_4$ and we introduce
an effective coupling $k$ defined by :
\begin{equation}
{\cal L}_{\gamma V W} = i \; k \;  \epsilon^{\mu \nu \alpha \beta} \;
\partial_\mu A_\nu \; (V_\alpha^+ W_\beta^- - V_\alpha^- W_\beta^+ ).
\end{equation}
The same type of coupling was derived in Ref. \cite{chivukula-golden}
for the $\omega_T Z \gamma$ interaction, where $\omega_T$ is the technicolor
analogue of the $\omega$-meson.
We would like to point out that there is no theoretical reason
to expect large values $|k| = {\cal O} (1)$. In fact,
$|k| \simeq 10^{-2}$ in QCD but in this paper we adopted the approach of
deriving what type of limits on $|k|$ could be achieved in future
$e \gamma$ colliders.

It is our purpose to study how one can constrain the possible values of $k$
by computing deviations from the Standard Model predictions. In this paper
we'll focus on the particular
process $e \gamma \rightarrow \nu_e W \gamma$, attainable at future $e \gamma$
colliders and
in the next section we compute the cross section for this process in the
anomalous BESS model.

\section{A possible signature for the anomalous BESS model in $e \gamma$
         colliders }

\hspace*{\parindent}

We are interested in the process  $e \gamma \rightarrow \nu_e W \gamma$,
depicted in Fig. 1.
The exact Standard Model cross section for this process including realistic
cuts was computed in Ref. \cite{kingman}.
We'll assume that the initial electron beam has
a monochromatic energy of $500$ GeV and that the initial photon has an
energy spectrum  originated
from  laser back-scattering off a $500$ GeV electron (positron) beam.
In this preliminary study we'll use the effective-$W$ approximation
\cite{effective-w} to
estimate the cross section and we won't try to implement any realistic
acceptance cuts. One can straightforwardly improve upon our results by
computing
an exact cross section with appropriate cuts for a more detailed study.
Since we work in unitary gauge we don't need to invoke the equivalence
theorem to derive our results.

The cross section can be written as :
\begin{equation}
\frac{d \sigma (s)}{d M_{\gamma W}} = \frac{2 M_{\gamma W}}{s} \; \;
\sum_{i=L,T} \;  \left( \frac{d {\cal L}}{d \tau} \right)_{\gamma W_i} \; \;
\hat{\sigma}_i (\tau s) ,
\end{equation}
where $M_{\gamma W}$ is the $\gamma - W$ invariant mass , $\hat{\sigma}_i$
is the subprocess $\gamma W \rightarrow \gamma W$ cross section for a given
initial $W$ helicity $i$,
$\tau = M^2_{\gamma W}/s$ and the so-called luminosity function
$d {\cal L}/d \tau$ is given by :
\begin{equation}
 \left( \frac{d {\cal L}}{d \tau} \right)_{\gamma W_i} =
\int_{\tau}^{x_m} \; \frac{d x}{x} \; f_{e \rightarrow \gamma} (x) \;
f_{e \rightarrow W_i} (\tau/x) .
\end{equation}
The functions $f_{e \rightarrow \gamma} (x)$ and
$f_{e \rightarrow W_i} (x)$ are respectively the energy spectrum of
the laser back-scattered photon and the probability of finding a $W$ of
polarization $i$ carrying an energy fraction $x$ of the parent electron.

The energy spectrum of a back-scattered photon as a function of
the energy
fraction $x$ of the initial electron's energy is given by \cite{laserback} :
\begin{equation}
f_{e \rightarrow \gamma} (x) = \frac{N(x,\xi)}{D(\xi)}
\end{equation}
where $\xi$ is defined in terms of the electron mass $m_{e}$, the electron
initial energy $E$ and the energy of the photon in the laser beam $\omega_0$ :
\begin{equation}
\xi = \frac{4 E \omega_0}{m_e^2}
\end{equation}
and
\begin{eqnarray}
 N(x,\xi) &=& 1 - x + \frac{1}{1-x} - \frac{4 x}{\xi (1-x)} +
                                       \frac{4 x^2}{\xi^2 (1-x)^2 }  \\
D(\xi) &=& \int_0^{x_m} \; dx \; N(x,\xi)   \nonumber \\
&=& \left[ 1 - \frac{4}{\xi} - \frac{8}{\xi^2} \right] \ln (1+\xi) +
\left(\frac{1}{2} + \frac{8}{\xi} - \frac{1}{2 (1+\xi)^2}  \right)
\end{eqnarray}
where $x_m = \frac{\xi}{1+\xi}$ is the maximum energy fraction
carried by the scattered photon. If $x_m \geq 0.828$ other processes start
to compete with the photon back-scattering reaction degrading the photon beam.
In our case, with a laser energy $\omega_0 = 1.17$ eV we have $x_m = 0.90$.
Therefore, here we'll take $x_m = 0.828$.

For $f_{e \rightarrow W_i} (x) $ we use \cite{effective-w} :
\begin{equation}
f_{e \rightarrow W_i} (x) = \left\{ \begin{array}{ll}
  \frac{\alpha}{8 \pi \sin^2 \theta} \; \frac{(x^2 + 2 (1-x))}{x} \;
\ln (\frac{4 E^2}{M_W^2})  & i = T  \\
  \frac{\alpha}{4 \pi \sin^2 \theta} \; \frac{1-x}{x}  & i = L
                                         \end{array}  \right.
\end{equation}

We now compute the point cross section that includes the diagrams shown in
Fig. 2, where the notation is defined. The amplitude is given by :
\begin{equation}
{\cal M} = {\cal M}_s + {\cal M}_t + {\cal M}_c
\label{eq:amplitude}
\end{equation}
where the subscripts $s,t$ and $c$ stand for the $s-$channel, $t-$chanel
and contact term contributions respectively.

We find :
\begin{eqnarray}
{\cal M}_s &=& \varepsilon_{1 \mu} \varepsilon_{2 \nu}
             \varepsilon_{3 \alpha}^\ast \varepsilon_{4 \beta}^\ast \left\{
\frac{i e^2}{(p_1 + p_2)^2 - M_W^2} \; \right.  \nonumber \\
& & \left[ (p_1 - p_2)^\lambda g^{\mu \nu} \; + \; 2 p_2^\mu g^{\nu \lambda}
\; - \; 2 p_1^\nu g^{\mu \lambda}  \right] \; g_{\lambda \sigma}
 \left[ (p_3 - p_4)^\sigma g^{\alpha \beta} \; + \; 2 p_4^\alpha
g^{\beta \sigma}
\; - \; 2 p_3^\beta g^{\alpha \sigma}  \right]  \nonumber \\
& & -\frac{i e^2 M_W^2}{(p_1 + p_2)^2 - M_W^2} \; g^{\mu \nu} g^{\alpha \beta}
\nonumber  \\
& & - \frac{i k^2}{(p_1 + p_2)^2 - M_V^2 + i M_V \Gamma_V} \;
\epsilon^{\lambda \mu \sigma \nu} \epsilon^{\lambda ' \alpha \sigma ' \beta}
p_{1 \lambda} p_{3 \lambda '} \left[ g_{\sigma \sigma '} -
 \left. \frac{Q_\sigma Q_{\sigma '} }{M_V^2} \right]  \right\} \nonumber \\
{\cal M}_t &=& {\cal M}_s  \;
(p_2 \leftrightarrow -p_4  \; , \;
\varepsilon_2  \leftrightarrow \varepsilon_4)  \\
{\cal M}_c &=& -i e^2  \varepsilon_{1 \mu} \varepsilon_{2 \nu}
             \varepsilon_{3 \alpha}^\ast \varepsilon_{4 \beta}^\ast
\left[ 2 g^{\mu \alpha} g^{\nu \beta} \; - \; g^{\mu \nu} g^{\alpha \beta}
\; - \;  g^{\mu \beta} g^{\nu \alpha} \right]
\end{eqnarray}

In order to compute the subprocess cross section $\hat{\sigma}_i (s)$
we square the amplitude Eq[\ref{eq:amplitude}], summing over the
helicities of all particles but
the initial $W$ and integrate it over all values of the scattering angle.

Before showing our results, let's recall that our free parameters are
$(M_V,\Gamma_V,k)$. In the limit $k \rightarrow 0$ one recovers
the Standard Model prediction. Here we'll concentrate on two sets of masses
 and widhts : $(M_V = 500 \;{\mbox{GeV}} \;,\; \Gamma_V =  50 \; {\mbox{GeV}})$
and  $(M_V = 800\; {\mbox{GeV}}\;,\; \Gamma_V =  100\; {\mbox{GeV}})$,
which are
chosen such that there is no unitarity violation in $W_L W_L$ scattering
\cite{Bagger2}.
For each set of $(M_V,\Gamma_V)$ we vary the values of $|k|$ from $0$ to
$0.3$. Our results for the subprocess cross section are displayed in
Figs. 3 and 4. Notice that the new resonance couples much more strongly
to the longitudinally polarized $W$'s, as expected from their Nambu-Goldstone
boson origin.

In Fig. 5  we present our results for the differential cross section
Eq. [39]. These plots follow simply from convoluting the subprocess cross
section with a rapidly falling luminosity function. In Table 1 we show
the the cross section in fentobarns integrated around the resonace peak
($M_V - \Gamma_V
\; < \; M_{\gamma W} \; < \; M_V + \Gamma_V$). For an $e \gamma$ collider
 with a total yearly luminosity of $100 \;
{\mbox{fb}}^{-1}$ \cite{vlepp} one expects to be able to put firm constraints
on this particular model.

We have also calculated the cross sections for $(M_V = 1000$ GeV, $\Gamma_V =
200$ GeV $)$ but due to the rapidly falling luminosity function we find that
these values are out of reach for a $1$ TeV $e^+e^-$ collider.

\section{Conclusion}
\hspace*{\parindent}

We have performed a preliminary study of the deviations from Standard Model
predictions of the cross section for the process
$e \gamma \rightarrow \nu_e \gamma W$ in the context of a higgsless chiral
lagrangian model that includes a new vector resonance $V$ and an anomalous
$\gamma W V$ interaction. The model has as free parameters the mass $M_V$,
the width $\Gamma_V$ and the
anomalous coupling $k$.
No exhaustive search for the discovery region in the $(M_V, \Gamma_V, k)$
parameter space was intended and no realistic cuts and experimental
efficiencies were included.

Our results suggest that if such a particle $V$ exists with a mass in the
range $500 - 800$ GeV  with an anomalous coupling strength $|k| \gsim
0.2$  one would expect an enhancement of ${\cal O} (50-100 \%)$
over the Standard Model result for the cross section
$e \gamma \rightarrow \nu_e \gamma W$ integrated around the new particle mass.
Even with a yearly luminosity of  $10 \;{\mbox{fb}}^{-1}$ our results imply
a number of events of the order of ${\cal O} (400 - 2000)$ per year,
which should not be difficult to find in the reasonable clean environment
of an $e \gamma$ collider. The situation improves for larger values of $|k|$
and a more
realistic analyses may be worthwile pursuing, where more definite constraints
on the parameter space of the model can be set once such a cross section
is measured.
For $|k| \lsim 0.1$, this process loses its sensitive to the presence of the
new resonance and one would not be able to distinguish between the model
studied here
and the Standard Model result.

An interesting question is the consequence of this model for future
hadron colliders, in particular in the mode $p p \rightarrow \gamma (Z) W X$.
Work along these lines is now in progress.

\section*{Acknowledgement}
\hspace*{\parindent}

This research was supported in part
by the National Science Foundation under Grant No. PHY-9001439, and by the
Texas National Research Laboratory Commission under Award No. RGFY9214.

\newpage
\section*{Figure Captions}
\hspace*{\parindent}

Figure 1 : The process $e \gamma \rightarrow \nu_e W \gamma$.

\bigskip

Figure 2 : Feynman diagrams for the subprocess $\gamma W \rightarrow
\gamma W$ cross section.

\bigskip

Figure 3a : Point cross section in fentobarns for the process
$W_L \gamma \rightarrow W \gamma $ as a function of the $\gamma
W$ center-of-mass energy with $M_V = 500$ GeV and $\Gamma_V = 50$ GeV.
Solid line: $|k| = 0$ (Standard Model);
dot-dashed line:$|k| = 0.1$ ; dotted line: $|k| = 0.2$;
dashed line: $|k| = 0.3$.

\bigskip

Figure 3b : Same as Figure 3a  but for the process $W_T \gamma \rightarrow
W \gamma$.

\bigskip

Figure 4a : Same as Figure 3a but for  $M_V = 800$ GeV and $\Gamma_V = 100$
GeV.

\bigskip

Figure 4b : Same as Figure 3b but for  $M_V = 800$ GeV and $\Gamma_V = 100$
GeV.

\bigskip

Figure 5a : Differential cross sections in fb/GeV for the process
$e \gamma \rightarrow \nu_e W \gamma$ as a function of the final state
$W \gamma$ invariant mass with $M_V = 500$ GeV and $\Gamma_V = 50$ GeV.
Solid line: $|k| = 0$ (Standard Model);
dot-dashed line:$|k| = 0.1$ ; dotted line: $|k| = 0.2$;
dashed line: $|k| = 0.3$.

\bigskip

Figure 5b : Same as Figure 5a but with  $M_V = 800$ GeV and $\Gamma_V = 100$
GeV.

\newpage
\begin{table}[h]
\begin{center}
{ \renewcommand{\arraystretch}{1.1}

\begin{tabular}{|c|c|c|}   \hline
\multicolumn{1}{|c|}{$|k|$}       &\multicolumn{1}{c|}{($500$ GeV, $50$ GeV)}
 &\multicolumn{1}{c|}{($800$ GeV, $100$ GeV)}   \\  \hline
$0$      &$128.1$  &$28.29$  \\  \hline
$0.1$    &$132.8$  &$29.12$  \\  \hline
$0.2$    &$202.4$  &$41.19$  \\  \hline
$0.3$    &$502.8$  &$93.32$  \\  \hline
$0.4$    &$1,311$  &$233.4$  \\  \hline
\end{tabular}   }
\end{center}
\vskip 0.1in

\caption{ Integrated cross section  around
$M_V - \Gamma_V < M_{\gamma W}
< M_V + \Gamma_V $ in fentobarns for different values of
$|k|$ and $(M_V,\Gamma_V)$.}

\end{table}

\clearpage

\end{document}